\documentclass[aps,prl,reprint,groupedaddress]{revtex4-2}

\usepackage{pdfpages} 
\usepackage{pgffor} 

\makeatletter
\AtBeginDocument{\let\LS@rot\@undefined}
\makeatother

\def\supplementfilename{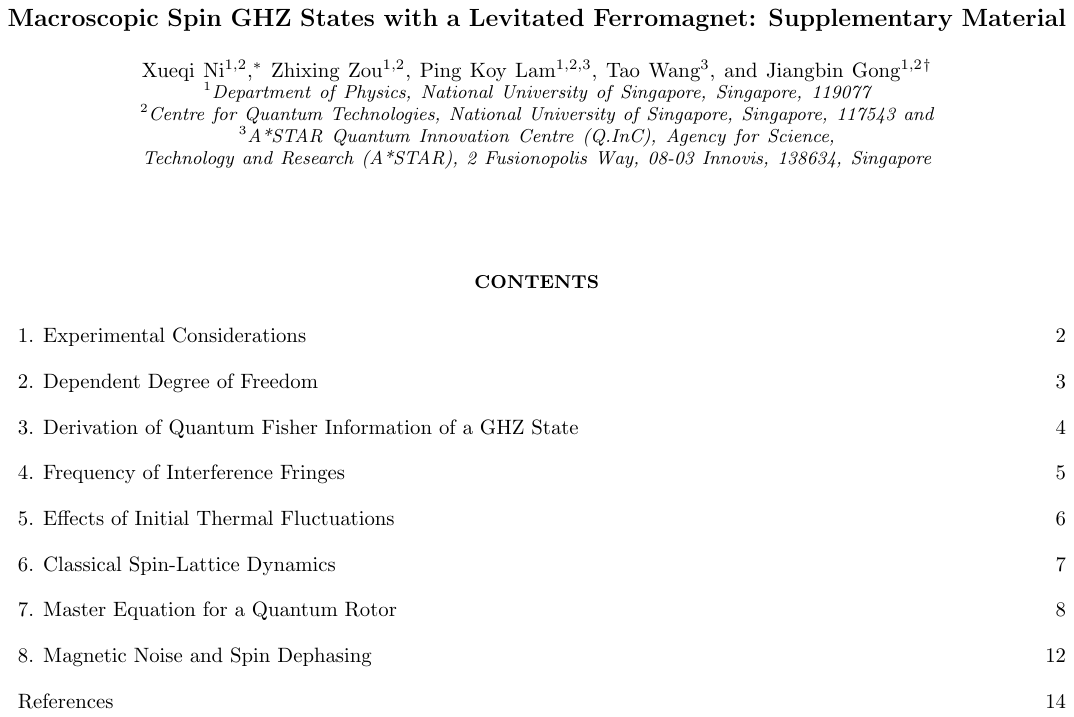}

\pdfximage{\supplementfilename}
\def\numbersupplementpages{\the\pdflastximagepages}

\newif\ifarXiv
\arXivtrue 

\usepackage{amsmath}
\usepackage{setspace} 
\usepackage{textcomp}
\usepackage{amsfonts}
\usepackage{amssymb}
\usepackage{graphicx}
\usepackage{siunitx}
\usepackage{setspace}
\usepackage{xcolor}
\usepackage{graphicx}
\usepackage{lipsum}
\usepackage{afterpage}
\usepackage[resetlabels,labeled]{multibib}
\newcites{BC}{BC Readings}
\usepackage[colorlinks, citecolor={blue}, linkcolor={black}]{hyperref}

\newcommand{\GJ}{\textcolor{black}}
\begin{document}


\title{\textbf{Macroscopic Spin GHZ States with a Levitated Ferromagnet} 
}%

\author{Xueqi Ni$^{1,2}$}
\email{Contact author: xueqi.ni@u.nus.edu}

\author{Zhixing Zou$^{1,2}$}

\author{Ping Koy Lam$^{1,2,3}$}
\author{Tao Wang$^{3}$}
\author{Jiangbin Gong$^{1,2}$}
\email{Contact author: phygj@nus.edu.sg}

\affiliation{$^1$Department of Physics, National University of Singapore, Singapore, 119077}
\affiliation{$^2$Centre for Quantum Technologies, National University of Singapore, Singapore, 117543}
\affiliation{$^3$A*STAR Quantum Innovation Centre (Q.InC), Agency for Science, Technology and Research (A*STAR), 2 Fusionopolis Way, 08-03 Innovis, 138634, Singapore}

\begin{abstract}
\GJ{The generation of macroscopic quantum states can drive both fundamental physics and quantum technologies}. This work proposes a top-down approach to the generation of macroscopic spin GHZ states using a levitated ferromagnet, where a strong locking between the collective spin and the lattice rotation enables mechanical control of the collective spin.  We quantify the metrological advantage of the resulting macrospin superposition state by showing that Heisenberg scaling of the quantum Fisher information is \GJ{achievable}. Roles of symmetry and geometry are analyzed in terms of decoherence due to gas collisions, identifying accessible conditions for experimental realization. The usefulness of a macrospin superposition state of a levitated cylindrical ferromagnet in testing spin-dependent wavefunction collapse models is also discussed. 
\end{abstract}

\maketitle










{\it Introduction.--} The Greenberger–Horne–Zeilinger (GHZ) state represents a paradigmatic form of multipartite entanglement\,\cite{greenbergerBellsTheoremInequalities1990} and provides a many-body analogue of Schrödinger’s cat\,\cite{leibfriedCreationSixatomSchrodinger2005a,omranGenerationManipulationSchrodinger2019,baoCreatingControllingGlobal2024}. In a spin ensemble, a GHZ state results from a coherent superposition \GJ{of two configurations depicting all the spins collectively aligned in two opposite directions\,\cite{zhangFastGenerationGHZlike2024a}, forming a macroscopic entangled} state\,\cite{frowisMacroscopicQuantumStates2018}. Such macroscopic quantum entanglement is of central importance to studies of the quantum-to-classical transition and to possible modifications to quantum mechanics for massive objects\,\cite{bassiModelsWavefunctionCollapse2013,boseMassiveQuantumSystems2025}. They also constitute a key resource for quantum sensing protocols that can surpass the standard quantum limit\,\cite{huelgaImprovementFrequencyStandards1997,jonesMagneticFieldSensing2009,matsuzakiMagneticFieldSensing2011,montenegroHeisenberglimitedSpinmechanicalGravimetry2025,braskImprovedQuantumMagnetometry2015,lvSupersensitiveNoiseSensing2025}. However, the generation of GHZ states with a large number of spins remains extremely challenging as it requires complex coherent control sequences \GJ{to be applied to multiple spins}~\cite{zhengOneStepSynthesisMultiatom2001,songGenerationMulticomponentAtomic2019,baoCreatingControllingGlobal2024}.    Recent bottom-up approaches to GHZ states build up entanglement from microscopic constituents such as high-spin nuclei\,\cite{yangMinutescaleSchrodingercatState2024,yuSchrodingerCatStates2025,guptaRobustMacroscopicSchrodingers2024}, yet \GJ{the effective total number} of spins therein remains much limited to date. 

 In a ferromagnet, strong exchange interactions between a large number of microscopic spins can cause them to behave collectively as a single macrospin\,\cite{heisenberg1928theorie,xiaoMacrospinModelsSpin2005,sayadMacrospinApproximationQuantum2012a,jacksonkimballPrecessingFerromagneticNeedle2016}. This macrospin is further coupled to the mechanical rotation of the lattice, enabling coherent conversion between spin and mechanical angular momentum\,\cite{niMicroscopicTheoryPrecessing2025b,sharmaSpinCatStates2021,kaniMagnomechanicalRotationalSchrodingers2025,heGenerationFourcomponentMagnonic2026}. By levitating the ferromagnet in high vacuum, clamping-induced dissipation is eliminated, and environmental interactions that lead to decoherence are significantly suppressed\,\cite{gonzalez-ballesteroLevitodynamicsLevitationControl2021,shengLevitatedMilligramscaleFerromagnetic2026,hoferHighMagneticLevitation2023a,fuwaFerromagneticLevitationHarmonic2023}. We hence propose in this work to use a levitated ferromagnet as an innovative top-down route toward the possible generation of macroscopic spin GHZ state via mechanical control. 
 
{In contrast to conventional spin-mechanical systems, where a single spin serves as an ancilla for generating motional quantum superposition\,\cite{yinLargeQuantumSuperpositions2013,scalaMatterWaveInterferometryLevitated2013a}, here the mechanical degree of freedom serves as a flexible control for coherently generating a spin GHZ-like state of the macrospin, with  the macrospin of a ferromagnet itself \GJ{accommodating} the macroscopic quantum resource and intrinsically encoding the entanglement among its constituent spins.} We further demonstrate that the quantum superposition state of the macrospin, i.e., a GHZ-like state, enables Heisenberg-limited magnetometry.  We also analyze the dominant decoherence channels to identify experimentally accessible regimes.  As a possible application, we advocate using the generated macrospin superposition state to test possible modifications to quantum mechanics where wavefunction collapse depends not only on spatial degrees of freedom, but also on spins\,\cite{bassiModelsWavefunctionCollapse2013}.

{\it Results.--}\GJ{Even before} Heisenberg proposed the exchange interaction between neighboring spins as the microscopic origin of ferromagnetism\,\cite{heisenberg1928theorie}, experiments such as the Einstein–de Haas effect and Barnett effect had revealed a deep interplay between magnetization and mechanical \GJ{degrees of freedom}\cite{einstein1915experimenteller,barnett1909magnetization,nieEinsteindeHaasEffect2025}. \GJ{Indeed, a previous microscopic study confirmed that in the single-domain regime of a ferromagnet, a collective magnetic moment arising from the constituent spins, hereafter denoted as $\hat{\mathbf{S}}$, can be strongly locked to the rigid-body rotation of the ferromagnet\,\cite{niMicroscopicTheoryPrecessing2025b}. Therefore, after averaging out the much faster internal vibrations}, a ferromagnet can behave as a single collective spin strongly coupled to a mechanical rotor that represents the collective lattice degree of freedom, with its orientation aligning with the magnetic anisotropy axis $\hat{\mathbf{n}} = (\sin\hat{\theta}\cos\hat{\phi},\sin\hat{\theta}\sin\hat{\phi},\cos\hat{\theta})$. The rotor motion is further characterized by three angular momentum components $(\hat{L}_x,\hat{L}_y,\hat{L}_z)$, and principal moments of inertia $(I_x,I_y,I_z)$. The spin-lattice coupling can then be \GJ{well described} by the nonlinear form as $-C(\hat{\mathbf{S}}\cdot \hat{\mathbf{n}})^2$, where the coupling coefficient $C>0$ determines the energy scale of the spin-lattice relaxation\,\cite{bandDynamicsMagneticNeedle2018}. This collective \GJ{spin-lattice interaction can make accurate predictions of the gyroscopic dynamics, including collective precession, nutation, and libration of the combined spin-lattice system\,\cite{niMicroscopicTheoryPrecessing2025b,bandDynamicsMagneticNeedle2018,fadeevFerromagneticGyroscopesTests2021,hdh6-r1gy} when benchmarked with a microscopic Hamiltonian theory\,\cite{niMicroscopicTheoryPrecessing2025b}. }


\begin{figure}[t]
    \centering
    \includegraphics[width=0.98\linewidth]{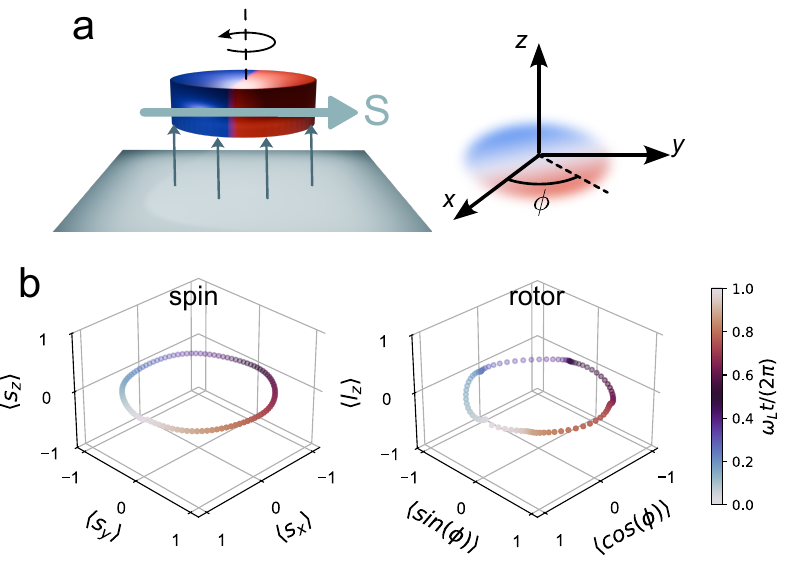}
    \caption{\textbf{Spin-lattice coupling in a levitated ferromagnet}. (a) Schematic of a levitated ferromagnet, where the collective spin couples to the mechanical rotation of the lattice. The orientation of the lattice is depicted by a rotation angle $\hat{\phi}$ in Cartesian coordinates. (b) Time evolution of the spin (left) and mechanical rotor (right) degrees of freedom in a static magnetic field, demonstrating locked precessional motion with finite-size dynamical fluctuations arising from spin-lattice angular-momentum exchange.}
    \label{fig:1}
\end{figure}

Below we focus on the precession dominant regime of a levitated ferromagnet, where the spin and lattice rotate about a fixed $z$-axis, as illustrated in Fig.~\ref{fig:1}(a). By neglecting rotations about the $x$ and $y$ axes, the three-dimensional rotor reduces to an effective planar rotor described by the conjugate variables $[\hat{\phi},\hat{L}_z] = i\hbar$, yielding the system Hamiltonian:
\begin{equation}
    H = \frac{\hat{L}_z^2}{2I_z} - C(\hat{\mathbf{S}}\cdot \hat{\mathbf{n}})^2.
\end{equation}
The collective spin $\hat{\mathbf{S}}$ spans a Hilbert space of dimension $(2N+1)$, corresponding to a total spin quantum number $S_0 = N\hbar$. As a demonstration of the corresponding collective quantum dynamics, Fig.~\ref{fig:1}(b) shows the evolution of the spin (left) and rotor (right) under a magnetic field of $B_z$ = 0.5 mT with $N = 250$. In our numerical calculations, both the spin and rotor angular momenta are expressed in dimensionless form as $\hat{\mathbf{s}} = \hat{\mathbf{S}}/S_0$ and $\hat{l}_z = \hat{L}_z/S_0$. The coupling coefficient $C$ is chosen as $\omega_L/(2N\hbar)$ throughout \GJ{this work} to ensure that the spin-lattice coupling energy remains extensive and scales linearly with $N$. The initial state for our dynamics studies is always chosen to be a product of coherent states of both the collective spin and the lattice rotor. During the evolution, the spin and lattice motions are \GJ{seen clearly from the presented numerics} to be synchronized or ``locked''.

The ``locked" dynamics between the collective spin and lattice orientation yield two primary physical implications. First, this strong coupling establishes a mechanical interface to drive the spin system into a non-classical regime. To achieve this, we break the rotational symmetry of the levitating landscape to create an angular double-well potential acting on the rotor (Fig.~\ref{fig:2}a; see Supplementary Material). The rotor experiences a dynamical instability at orientation $\hat{\phi}=0$, which is a saddle point, carrying the rotor into a superposition of clockwise and counterclockwise rotation\,\cite{maQuantumMetrologyFloquetengineered2025}. At short times, the dynamics can be understood from a rotational inverted harmonic oscillator described by the Hamiltonian $H = \hat{L}_z^2/{2I_z} - I_z\omega^2 \hat{\phi}^2/2$. The underlying instability leads to an exponential growth of angular uncertainty, $ \Delta \hat{\phi}(t) \propto \Delta \hat{\phi}(0) \exp(\omega t) \propto N^{-1/2}\exp(\omega t)$. As a result, the transition from quasi-classical to fully quantum dynamics occurs on a remarkably short logarithmic timescale\,\cite{roda-llordesMacroscopicQuantumSuperpositions2024,zhouSpindependentForceInverted2025}. This is in contrast to massive systems undergoing regular motion, where the emergence of quantum effects occurs over much longer polynomial timescales\,\cite{MVBerry1979, BERMAN1981}.   Fig.~\ref{fig:2}b shows the numerically obtained delocalization time $t_E$ with total spin number $N$, under an angular potential in the form of $V(\hat{\phi}) = V_1 \cos{\hat{\phi}} + V_2 \cos^2{\hat{\phi}} + V_3\cos^3{\hat{\phi}}$, where coefficients $V_i$ control the barrier height and asymmetry of the potential. The timescale follows a scaling law of $t_E \approx 2.48 \ln N$ ns, implying that even for a system with a total spin number $N = 10^{7}$, macroscopic quantumness emerges within 50 ns, a short timescale favorable for beating decoherence.

\begin{figure*}
    \centering
    \includegraphics[width=0.98\linewidth]{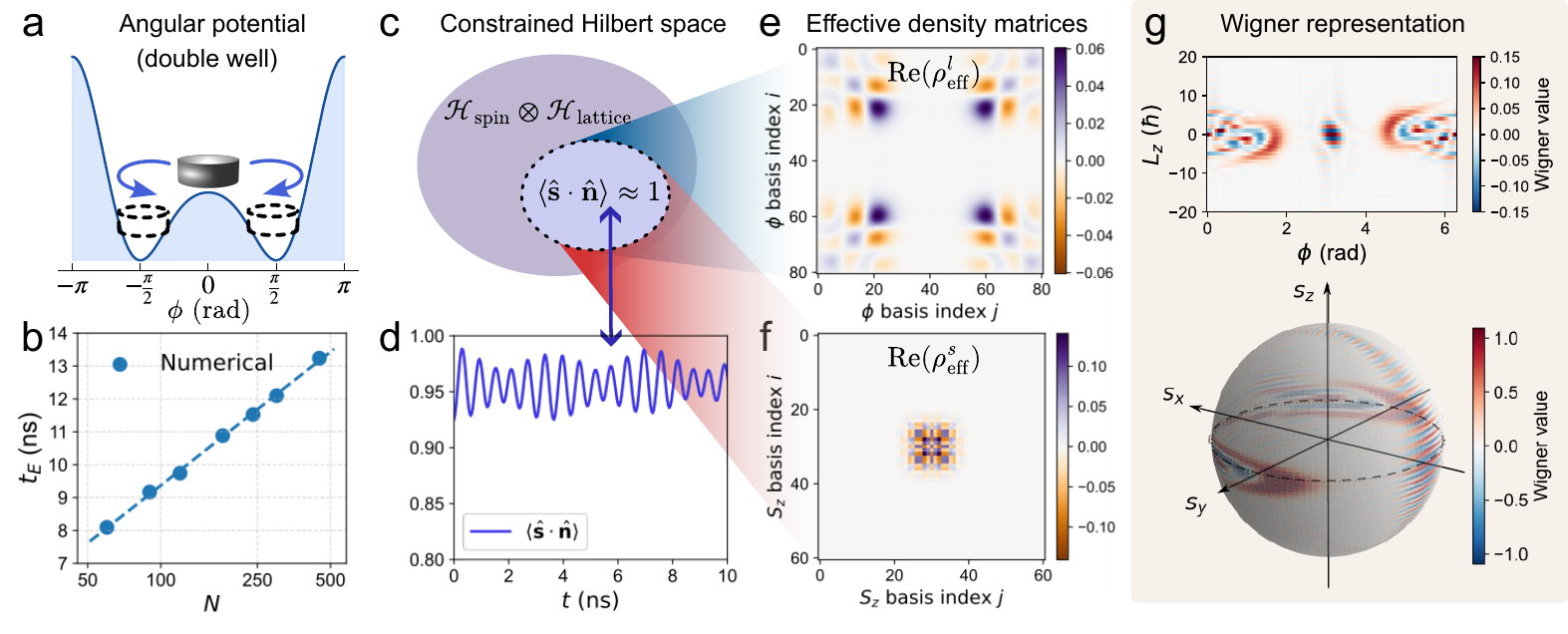}
    \caption{\textbf{Generation and phase space representation of macrospin GHZ states.} (a) Illustration of a levitated ferromagnet in an angular double-well potential. (b) Delocalization time $t_E$ versus spin number $N$. Numerical simulation results (dots) follow a logarithmic dependence fitted by $t_E \approx 2.48 \ln N$ (dashed line), using parameters $(V_1,V_2,V_3)/N\hbar\omega_L = (-0.015,0.05,-0.025)$. (c) Conceptual mapping of the full tensor-product Hilbert space onto the constrained subspace. (d) $\langle \hat{\mathbf{s}}\cdot\hat{\mathbf{n}} \rangle$ vs time, reflecting strong spin-lattice locking. (e) Real part of effective density matrix $\hat{\rho}_{\rm eff}^{l}$ in the rotor basis. Basis indices are ordered from 0 to $2\pi$. (f) Real part of the effective density matrix $\hat{\rho}_{\rm eff}^{s}$ in the spin $\hat{S}_z$ basis. (g) Wigner distribution of the rotor (top) and spin (bottom) obtained from $\hat{\rho}_{\rm eff}^{l}$ and $\hat{\rho}_{\rm eff}^{s}$.}
    \label{fig:2}
\end{figure*} 

During the rapid delocalization of the wavefunction, the spin-lattice locking remains robust. To quantify this locking, we introduce a correlator $\hat{\mathbf{s}}\cdot\hat{\mathbf{n}}$, which measures the alignment between the collective spin and the lattice axis. For a finite number of spins, \GJ{dynamical fluctuations} prevent perfect alignment, leading to $\langle \hat{\mathbf{s}}\cdot\hat{\mathbf{n}} \rangle < 1$, whereas the limit $\langle \hat{\mathbf{s}}\cdot\hat{\mathbf{n}} \rangle \to 1$ emerges as $N \to \infty$. For the levitated ferromagnet evolving under the angular potential, we find that $\langle \hat{\mathbf{s}}\cdot\hat{\mathbf{n}} \rangle \approx 1$ throughout the dynamics (Fig.~\ref{fig:1}d), indicating strong spin–lattice locking even in the presence of \GJ{exponential instability in the precessing dynamics}. We further observe small oscillations in the correlator, originating from the finite lattice inertia that limits the speed of angular-momentum transfer and reflects the backaction between spin and lattice motion.

A second implication of the spin-lattice locking condition is to enforce a physical mechanism for reducing the system’s independent degrees of freedom. This behavior is conceptually analogous to symmetry-induced entanglement in indistinguishable particles\,\cite{sasakiEntanglementIndistinguishableParticles2011,reuschEntanglementWitnessesIndistinguishable2015}, in which the wavefunction is restricted to a specific symmetric or antisymmetric form rather than the full tensor-product Hilbert space of subsystems. Here, the spin-lattice coupling similarly constrains the dynamics to a subspace $\mathcal{H}_c \subset \mathcal{H}_{\text{spin}} \otimes \mathcal{H}_{\text{lattice}}$ defined by the strong locking condition $\langle \hat{\mathbf{s}}\cdot\hat{\mathbf{n}} \rangle \approx 1$ (Fig.~\ref{fig:2}c). A constraint-adapted orthonormal basis for this constrained Hilbert space can be written as $\lvert \phi\rangle \otimes \lvert S_0;\mathbf{n}(\phi)\rangle$, where $|\phi\rangle$ denotes a rotor eigenstate with orientation angle $\phi$, and $ \lvert S_0;\mathbf{n}(\phi)\rangle = e^{-i\phi \hat{S}_z/\hbar}|S_x\rangle$ is the corresponding spin coherent state rotated to align with the lattice orientation.

Because of \GJ{the above-highlighted} physical constraint, the quantumness of the subsystems (spin or lattice) cannot be captured by a standard partial trace, which treats the two degrees of freedom as independent. \GJ{That is, we can exclude any severe destructive operations such as a measurement scheme that can break the spin-lattice locking}. Then, physical observables always need to be evaluated within the constrained Hilbert space. We therefore define the rotor Wigner quasiprobability distribution directly in this subspace as
\begin{equation}
\begin{aligned}
    W_{\rm rotor}(L_z,\phi)
    &=
    \mathrm{Tr}_{\mathcal{H}_{\rm c}}
    \!\left[
        \hat{\rho}\,
        \hat{W}_L(L_z,\phi)\otimes \hat{I}_s
    \right] \\
    &=
    \int_0^{2\pi} d\phi_1\,
    \langle \phi_1\rvert\,
    \hat{\rho}_{\rm eff}^l\,
    \hat{W}_L(L_z, \phi)\,
    \lvert \phi_1\rangle ,
\end{aligned}
\label{rhodefine}
\end{equation}
where $\hat{W}_L$ is the conventional rotor Wigner-Weyl transform operator acting only on the lattice degree of freedom\,\cite{kastrupWignerFunctionsPair2016}, $\hat{I}_s$ is the identity operator with the dimension of spin, and $\hat{\rho}$ is the original full density matrix. \GJ{In Eq.~(\ref{rhodefine}) above} we have introduced an effective density matrix for the rotor part, defined as
\begin{equation}
    \langle \phi_1\rvert
    \hat{\rho}_{\rm eff}^l
    \lvert \phi_2\rangle
    =
    \langle S;\mathbf{n}(\phi_1)\rvert
    \otimes
    \langle \phi_1\rvert\,
    \hat{\rho}\,
    \lvert \phi_2\rangle
    \otimes
    \lvert S;\mathbf{n}(\phi_2)\rangle .
\end{equation}
which maps the constrained spin-rotor system onto an effective density matrix in the pure rotor Hilbert space. This construction allows the quantum coherence of the macroscopic spin-rotor system to be characterized without artificially discarding correlations imposed by the physical constraints. A similar effective density marix $\hat{\rho}_{\rm eff}^s$ for the macrospin part can be constructed for the spin degree of freedom (Supplementary Material).

With the spin-lattice locking fully accounted for as a  physical constraint, we are now ready to investigate quantum coherence via individual phase spaces (Figs.~\ref{fig:2}e–g). The real part of $\hat{\rho}_{\rm eff}^l$ in Fig.~\ref{fig:2}e displays a characteristic quantum superposition structure. The dominant diagonal terms at basis indices associated with $\phi \approx \pi/2$ and $3\pi/2$ represent the two macroscopically distinct orientations of the rotor, whereas the symmetric off-diagonal elements confirm the persistence of quantum coherence. Such coherence is further reflected in the rotor Wigner distribution (Fig.~\ref{fig:2}g, top) as rapid interference fringes around the unstable point $\phi = \pi$, strong evidence of a Schrödinger's cat state. Due to the robust spin-lattice coupling, this non-classicality in the rotor degree of freedom is \GJ{equally manifested in} the magnetic part. The spin Wigner distribution (Fig.~\ref{fig:2}g, bottom) concurrently develops two antipodal maxima at $(\langle \hat{s}_x\rangle, \langle \hat{s}_y\rangle, \langle \hat{s}_z\rangle) = (0, \pm 1, 0)$, bridged by fast oscillations and negative regions, indicating the underlying macroscopic quantum coherence of the collective spin. This phase-space structure is consistent with a macroscopic GHZ state of the form $(\lvert+\rangle^{\otimes N} + \lvert-\rangle^{\otimes N})/\sqrt{2}$, where $\lvert+\rangle$ and $\lvert - \rangle$ are the two eigenstates of the spin-$y$ operator $\hat{\sigma}_y$. Such macrospin GHZ-like behavior  \GJ{reflects a crucial feature of the coupled spin-lattice system here: the macroscopic quantumness is not confined to a single degree of freedom, but is intrinsically shared across two strongly coupled degrees of freedom of a single massive object.}

\begin{figure}
    \centering
    \includegraphics[width=0.98\linewidth]{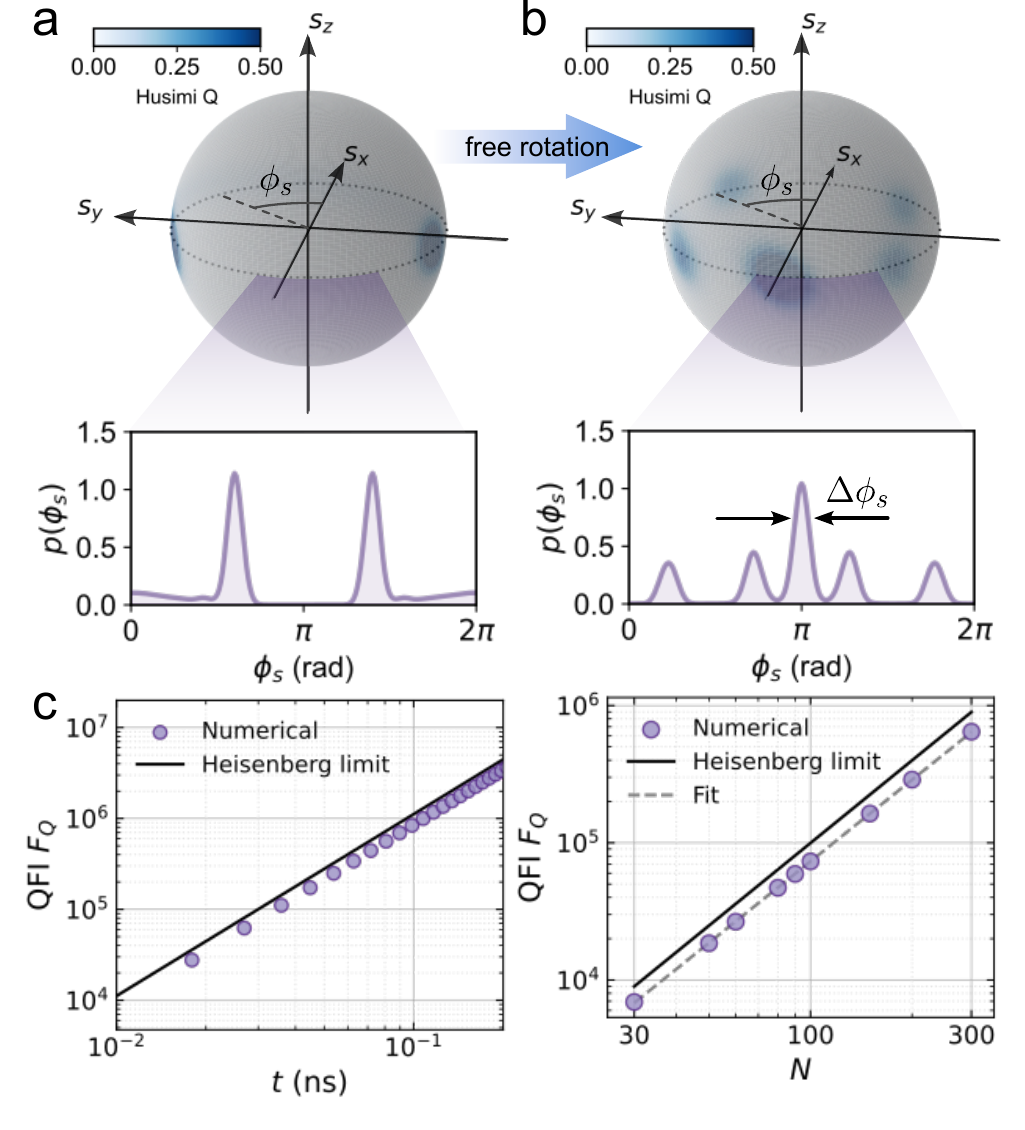}
    \caption{\textbf{Readout protocol and Heisenberg-limited magnetometry.} (a-b) Spin Husimi Q distributions representing the recombination of superimposed states via free rotation, and corresponding probability distributions of magnetization angle $\phi_s$ (a) before and (b) after recombination. (c) Quantum Fisher information for a transverse magnetic field $B_y$ as a function of time (left) and number of spins (right).}
    \label{fig:3}
\end{figure}

To experimentally read out the quantum coherence of the macroscopic GHZ state thus generated, we propose an intrinsic recombination protocol in which the mechanical rotation acts as the interferometer, but the coherence is detected through spins. Specifically, after the external double-well potential is switched off, the angular momentum stored in the lattice causes it to rotate freely. Through spin-lattice locking, this free mechanical rotation drives a coherent evolution of the collective spin state. As a result, the two macroscopically distinct GHZ components, $\lvert +\rangle^{\otimes N}$ and $\lvert -\rangle^{\otimes N}$, are dynamically brought into overlap in phase space. Their relative phases are converted into oscillatory interference patterns along the equator of the Bloch sphere. Fig.~\ref{fig:3}a and b demonstrate this process through the Husimi Q-function, a probability distribution in phase space, and its corresponding map onto the magnetization angle $\phi_s = \arctan(s_y/s_x)$. The frequency of the interference fringes $2\pi/\Delta\phi_s$ scales linearly with $N$ (Supplementary Material), reflecting the collective phase sensitivity of the GHZ state. The protocol realizes a mechanically induced interferometer with magnetization-based readout, providing a direct measurable signature of the GHZ-like quantum coherence in the levitated ferromagnets. {It should be noted that other witnesses of quantumness in uniformly precessing spin systems may be considered as well, where observables such as $S_x$ are associated with a score function admitting a classical bound that can be maximally violated by a GHZ state\,\cite{chenEvenparityPrecessionProtocol2024,tsirelsonHowOftenCoordinate2006}.}



The generation and interference readout of a macrospin superposition state enable the ferromagnet as a high-precision quantum sensor, providing opportunities to enhance the sensitivity of existing levitated sensors\,\cite{vancampAccurateTransferFunction2000}. The achievable sensitivity for magnetic field sensing can be quantified by the quantum Fisher information (QFI) $F_Q$\,\cite{huangEntanglementenhancedQuantumMetrology2024}. Consider our macroscopic GHZ state in the presence of a transverse magnetic field $B_y$, the field amplitude is encoded into the relative phase accumulated during evolution:
\begin{equation}
\lvert\psi (t)\rangle = \frac{1}{\sqrt{2}}\left(\lvert-\rangle^{\otimes N} + e^{2 i \gamma B_y N t/\hbar}\lvert +\rangle^{\otimes N}\right).
\end{equation}
The quantum Cramér-Rao bound predicts the minimum uncertainty in magnetic-field estimation as $\Delta B_y \ge 1/\sqrt{F_Q}$, where the corresponding QFI is $F_Q = (2\gamma N t)^2$ (Supplementary Material). This scaling is also known as the Heisenberg limit with respect to the number of spins $N$. Applying this framework to our macrospin superposition state, we demonstrate near-Heisenberg-limited scaling as a function of evolution time $t$ and the number of spins $N$, as shown in Fig.~\ref{fig:3}c. Our numerical results follow $F_Q \approx 0.83\,(2\gamma N t)^2$, where the prefactor $0.83$ quantifies the fidelity between the generated state and the ideal GHZ state. With Heisenberg-limited magnetometric performance, a magnetic field uncertainty down to $10^{-19}\,\mathrm{T}\cdot t^{-1}$ can be realized with $10^{7}$ spins, opening a pathway toward detecting ultra-weak magnetic signals relevant to searches for axion dark matter and other exotic spin-dependent interactions\,\cite{bassQuantumSensingParticle2024}. The achievable number of $N$ is practically limited by ambient magnetic field noise that causes dephasing of the GHZ state (Supplementary Material).

\begin{figure}
    \centering
    \includegraphics[width=0.98\linewidth]{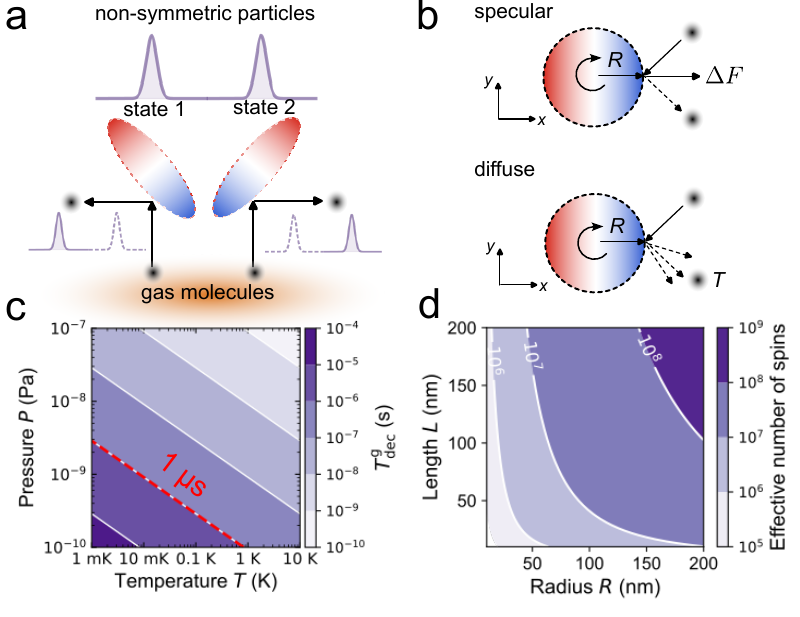}
    \caption{\textbf{Gas collisions and decoherence time scales.} (a) For a non-symmetric object, a single gas-molecule collision can reveal the underlying orientation and thus induce rapid decoherence. (b) \GJ{Schematics of specular and diffuse collisions, as explained in the main text}. (c) Required pressures and temperatures corresponding to different decoherence time scales due to diffuse gas collisions for a cylinder with aspect ratio $R/L = 0.5$ and $N = 10^7$, surrounded by $\mathrm{N}_2$ gas molecules. The dashed line corresponds to a 1 $\mu$s coherence window. (d) Effective spin numbers for a cylindrical nanomagnet calculated using $N = M_s\pi R^2L/g\mu_B$, where $M_s=1.44\times 10^6$ A/m is the saturated magnetization, $g$ is the Landé $g$-factor, and $\mu_B$ is the Bohr magneton constant.}
    \label{fig:4}
\end{figure}

{\it Effects of decoherence.--}\GJ{We now turn from the above ideal dynamics to more realistic situations with decoherence caused by gas collisions}.  Although the GHZ state generation discussed above is largely independent of the specific ferromagnet geometry, decoherence due to gas collision can depend strongly on symmetry and geometry. The symmetry of a levitated object determines whether the colliding gas molecules can reveal `which-orientation' information (Fig.~\ref{fig:4}a). For an asymmetric rotor, even a single collision can act as a strong measurement of its orientation, thereby collapsing the wavefunction and reducing the system to a classical mixture. This decoherence channel can be strongly suppressed if the levitated object possesses rotational symmetry, such as a sphere or a cylinder rotating about its symmetry axis. 

For a symmetric rotor, its geometry also controls the specific decoherence rate associated with angular-momentum diffusion. Two limiting regimes can be distinguished: specular and diffuse collisions\,\cite{qian2023generalized}. For specular collisions, gas molecules are reflected elastically from the surface of the levitated object. The resulting momentum transfer is normal to the local surface and therefore predominantly drives translational motion rather than rotational diffusion (Fig.~\ref{fig:4}b). In the diffuse collision regime, the outgoing molecular momenta are randomized through thermalization with the particle surface and follow a Maxwell-Boltzmann distribution at temperature $T$. We can consider the gas environment as a thermal bath and model this process using a Caldeira-Leggett type master equation (See Ref.\,\cite{Stickler2018} or Supplementary Material). The associated decoherence rate is given by $2D_\phi/\hbar^2$, where $D_\phi$ is the classical rotational diffusion constant, which quantifies the rate at which random thermal gas collisions inject angular momentum noise into the rotor. For a cylinder rotating around its symmetry axis, the diffusion constant is given by\,\cite{martinetzGasinducedFrictionDiffusion2018b}
\begin{equation}
    D_\phi = \frac{1}{2}PR^3L\sqrt{2\pi m k_B T} (2+R/L),
\end{equation}
where $R$ and $L$ are the radius and height of the cylinder, respectively, which together determine the effective collision surface; $P$ is the pressure of the environment, which sets the collision rate; and $m$ is the mass of the gas molecule. Given a fixed aspect ratio $R/L = 0.5$, the corresponding decoherence timescale $T_{\rm{dec}}^{\rm{g}} = \hbar^2/2D_\phi$ decreases with spin number approximately as $N^{-4/3}$. For a coherence window of approximately $1~\mu s$ and an effective spin number up to $10^{7}$, the required pressure needs to be down to $10^{-9}$ Pa, which is in the ultrahigh vacuum regime with a cryogenic temperature of $T = 10$ mK (Fig.~\ref{fig:4}c). The corresponding size of the ferromagnet is around 120 nm, placing it in the single-domain regime while being far from superparamagnetism (Fig.~\ref{fig:4}d)\,\cite{brownjr.FundamentalTheoremTheory1969}. A further discussion of thermal fluctuations affecting the rotor's initial state is in the Supplementary Section 5.

{\it Discussions and Conclusions.--} \GJ{A macroscopic GHZ state can be used for the test of possible modifications to quantum mechanics through the so-called spontaneous collapse models\,\cite{bassiModelsWavefunctionCollapse2013,schrinskiCollapseinducedOrientationalLocalization2017a,altamuraImprovedBoundsCollapse2025}. A key advantage of a rotationally symmetric ferromagnet used here is that collapse mechanisms based on position or mass density alone are strongly suppressed because their coupling to external sources is isotropic. Consequently, the orientational coherence is largely immune to conventional space- or gravity-based channels. 
The two superposition branches in the generated GHZ state can hence be indistinguishable in space, yet remain distinguishable through the internal spins, thus providing a particularly clean platform for testing collapse models exclusively due to the spin degree of freedom\,\cite{bassiNumericalAnalysisSpontaneous2004,pearleCosmogenesisCollapse2012}. }

\GJ{We have shown that a levitated ferromagnet offers a distinctive route toward macroscopic spin GHZ states by combining two strongly coupled degrees of freedom, spin and lattice, within a single massive object. The coupled degrees of freedom create new ways to generate, detect, and interpret macroscopic quantum superpositions.  {The use of a rotationally symmetric planar ferromagnetic rotor to accommodate macrospin superposition states is particularly appealing for both coherence protection and fundamental tests.} Looking ahead, it will be interesting to explore whether we can further stabilize such macroscopic coherence through periodic driving\,\cite{baiFloquetEngineeringOvercome2023,wachterGyroscopicallyStabilizedQuantum2026}.  We conclude that levitated ferromagnets can offer a unique testbed for macroscopic quantum physics, precision sensing, and possible modifications to quantum mechanics.}

\section{Acknowledgment} We thank Steven Touzard and Valerio Scarani for stimulating discussions. We acknowledge the
support of the National Research Foundation, Singapore,
under its Competitive Research Programme (CRP Award
No. NRF-CRP30-2023-0002). J.G. also acknowledges support from the National Research Foundation, Singapore, through the National Quantum Office, hosted in A*STAR, under its Centre for Quantum Technologies Funding Initiative (S24Q2d0009).

\bibliography{reference}
\label{LastBibItem}
\ifarXiv
    

\section*{Supplementary material}

## 1. EXPERIMENTAL CONSIDERATIONS

The crucial part of generating macrospin GHZ state in experiments is the preparation of an angular double-well potential and the control of the levitated objects. In principle, an angular double-well potential can be achieved by breaking the rotational symmetry of the levitation environment, leading to anisotropic coupling, as illustrated in Fig. S1.

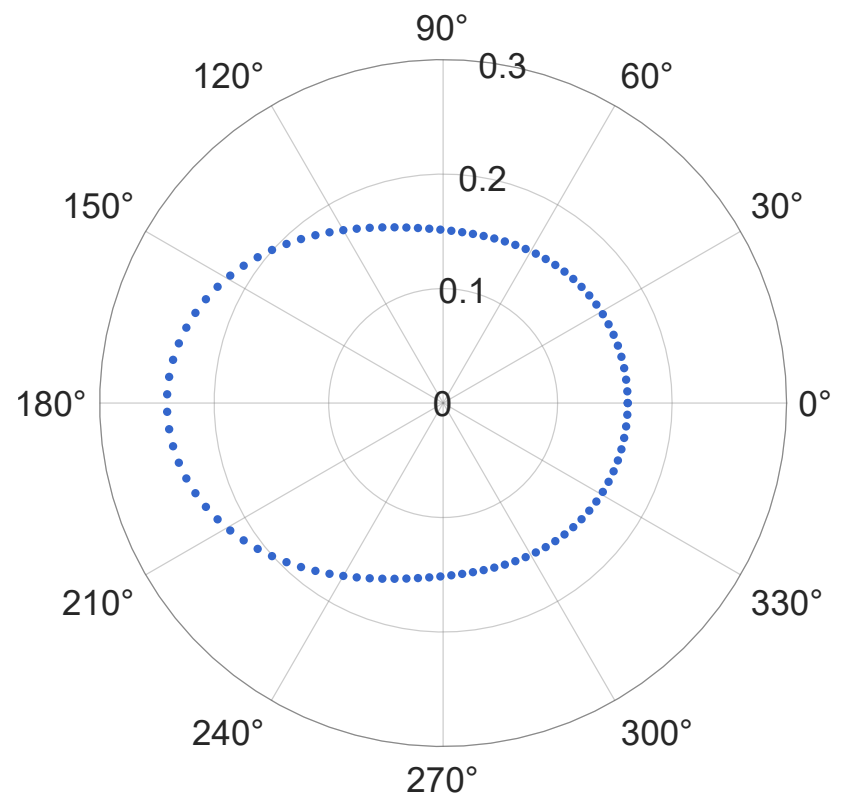


FIG. S1: Double-well potential $V(\phi)$ in the angular coordinate $\phi$, displayed as a polar plot. Symmetry breaking of the levitation environment lifts the rotational degeneracy and produces two degenerate energy minima at approximately $90^\circ$ and $270^\circ$, separated by a potential barrier.

The angular double-well potential is similar to the saddle potential typically used in the Paul trap [1]. Several experimental approaches can create such a potential:

*Anisotropic superconductor*:

In a type-II superconducting substrate, crystallographic defects and twin boundaries oriented along a preferred in-plane axis break the azimuthal symmetry of the Meissner–pinning landscape, producing an angular potential with two degenerate minima. The barrier height can be tuned via the levitation height, or engineered through directed ion irradiation and patterned superconducting thin films. The main challenge is reproducibility: naturally occurring defects are sample-dependent, while engineered defect landscapes add fabrication complexity and may introduce vortex-motion noise that dephases the macrospin.

*Magnetic bias field and hybrid superconducting–magnetic trapping scheme*:

A particularly promising alternative is the hybrid superconducting-magnetic trapping scheme. In this approach, the levitated object is supported by a superconducting substrate, which provides a stable and low-dissipation trapping environment. A weak in-plane magnetic field $B_{\text{bias}}$, applied via external coils, can break the rotational symmetry of the trap and generate an orientation-dependent potential through anisotropic Zeeman coupling. By combining multiple coils or employing shaped current distributions (e.g., quadrupole or higher-order configurations), one can engineer angular potentials that, in conjunction with the trapping environment, form a symmetric double-well with tunable barrier height and controllable well orientation. This approach offers high flexibility and real-time control. This scheme effectively separates stability (set by the superconducting levitation) from tunability (set by the magnetic control), enabling both long coherence times and flexible manipulation of the potential landscape. The remaining challenge is again technical noise from the coils, which necessitates highly stable current sources, low-noise electronics, and careful engineering to ensure compatibility with the superconducting environment.

*Optical torque:*

A particle with finite anisotropic optical polarizability in an elliptically polarized optical tweezer experiences an orientation-dependent optical potential. By superposing two orthogonally polarized beams with independently controlled intensities, one may realize a symmetric angular double-well with continuously tunable barrier height. The key constraint is that this approach requires a particle with intrinsic optical anisotropy, such as a sphere with an embedded dipole axis. Furthermore, photon scattering from the trapping field introduces an additional decoherence channel that needs to be mitigated, for instance, by using long-wavelength trapping light or cavity-assisted cooling.

## 2. DEPENDENT DEGREE OF FREEDOM

The entanglement between the collective spin and the rotor originates from their intrinsic spin-lattice coupling and is therefore fundamentally constrained by the underlying physics. As a consequence, the dynamically accessible Hilbert space is not the full tensor product $\mathcal{H}_{\text{spin}} \otimes \mathcal{H}_{\text{lattice}}$, but rather a restricted subspace selected by the strong spin-lattice locking condition $\langle \hat{\mathbf{s}} \cdot \hat{\mathbf{n}} \rangle \approx 1$.

Under this locking condition, for a given rotor orientation angle $\phi$, the spin is confined to a unique orientation determined by $\phi$. This induces a mapping from the full spin Hilbert space $\mathcal{H}_{\text{spin}}$ onto a one-dimensional subspace spanned by the rotated state

$$|S; \mathbf{n}(\phi)\rangle = \exp(-i\phi \hat{S}_z/\hbar)\, |S_x\rangle, \tag{1}$$

where $|S_x\rangle$ denotes the maximal eigenstate of $\hat{S}_x$. As a result, the constrained Hilbert space $\mathcal{H}_{\text{c}}$ has the same dimensionality as $\mathcal{H}_{\text{lattice}}$, and the accessible quantum states are effectively characterized by a single degree of freedom.

A convenient orthogonal basis for $\mathcal{H}_{\text{c}}$ is given by

$$|\phi\rangle \otimes |S; \mathbf{n}(\phi)\rangle, \tag{2}$$

in which the spin and rotor are not independent subsystems but rather dependent degrees of freedom tied together by the locking constraint. In this situation, it is unphysical to trace out one component as if it were an independent subsystem.

We therefore define the phase-space Wigner quasiprobability distributions over the constrained Hilbert space, where the operator $\hat{W}$ is the traditional Wigner-Weyl transform but the trace is over $\mathcal{H}_c$:

$$W_{\text{rotor}}(L_z, \phi) = \text{Tr}_{\mathcal{H}_{\text{c}}} \left[\hat{\rho}\, \hat{W}_L(L_z, \phi) \otimes I_s\right], \tag{3}$$

$$W_{\text{spin}}(\Theta, \Phi) = \text{Tr}_{\mathcal{H}_{\text{c}}} \left[\hat{\rho}\, I_L \otimes \hat{W}_S(\Theta, \Phi)\right]. \tag{4}$$

Specifically, when expanding the partial trace, the rotor Wigner function can be written as

$$\begin{aligned} W_{\text{rotor}}(L_z, \phi) &= \text{Tr}_{\mathcal{H}_{\text{c}}} \left[\hat{\rho}\, \hat{W}_L(L_z, \phi) \otimes I_s\right] \\ &= \int_0^{2\pi} d\phi_1\, \langle S; \mathbf{n}(\phi_1)| \otimes \langle \phi_1|\, \hat{\rho}\, \hat{W}_L \otimes I_s\, |\phi_1\rangle \otimes |S; \mathbf{n}(\phi_1)\rangle \\ &= \int_0^{2\pi} d\phi_1\, \langle \phi_1|\, \hat{\rho}^l_{\text{eff}}\, \hat{W}_L\, |\phi_1\rangle, \end{aligned} \tag{5}$$

where we have introduced an effective density matrix defined by

$$\langle \phi_1|\hat{\rho}^l_{\text{eff}}|\phi_2\rangle = \langle S; \mathbf{n}(\phi_1)| \otimes \langle \phi_1|\, \hat{\rho}\, |\phi_2\rangle \otimes |S; \mathbf{n}(\phi_2)\rangle. \tag{6}$$

This effective density matrix provides a complete description of the quantum state within the constrained Hilbert space. A complete construction in the spin basis is generally problematic, because the constrained subspace $\mathcal{H}_c$ typically exceeds the spin Hilbert space $\mathcal{H}_{\text{spin}}$ in dimension, and spin coherent states corresponding to different orientations are non-orthogonal at finite $N$. Only in the limit $N \to \infty$ do these orientations become orthonormal, allowing such a construction to be recovered.

Nevertheless, we can still construct an effective density matrix in the spin basis by performing an inverse Wigner-Weyl transformation to $W_{\text{spin}}$. When $N$ is large, where spin coherent states associated with different orientations become approximately orthogonal, this construction acquires a simpler basis representation as

$$\begin{aligned} \langle m_1|\hat{\rho}^s_{\text{eff}}|m_2\rangle &= \sum_i \sum_j (c_i^{m_1})^* c_j^{m_2} \langle S; \mathbf{n}(\phi_i)| \otimes \langle \phi_i|\, \hat{\rho}\, |\phi_j\rangle \otimes |S; \mathbf{n}(\phi_j)\rangle \\ &= \sum_i \sum_j (c_i^{m_1})^* c_j^{m_2} \rho^{ij}_{\text{eff}}. \end{aligned} \tag{7}$$

Here $|m\rangle$ denotes an eigenstate of $\hat{S}_z$. In the large-$S$ limit, the family of locked spin states $|S;\mathbf{n}(\phi_i)\rangle$ may be used as an approximately orthogonal basis, so that $|m\rangle$ can be expanded as a superposition of these states

$$|m\rangle = \sum_i c_i^m |S;\mathbf{n}(\phi_i)\rangle = \sum_i c_i^m \exp(-i\phi_i S_z/\hbar)|S_x\rangle, \tag{8}$$

with the coefficients $c_i^m$ given by

$$c_i^m = \langle S;\mathbf{n}(\phi_i)|m\rangle = \frac{\exp(im\phi_i)}{2^S}\sqrt{\binom{2S}{S+m}}. \tag{9}$$

## 3. DERIVATION OF QUANTUM FISHER INFORMATION OF A GHZ STATE

In estimation theory, the Fisher information sets the fundamental bound on the precision of parameter estimation. For an unbiased estimator, the measurement precision is bounded by the Cramér–Rao inequality,

$$\Delta\theta \geq \frac{1}{\sqrt{F_C(\theta)}}, \tag{10}$$

where the classical Fisher information is defined as

$$F_C(\theta) = \sum_i P(x_i|\theta)\left(\frac{\partial \ln P(x_i|\theta)}{\partial\theta}\right)^2. \tag{11}$$

Here, $P(x_i|\theta)$ is the conditional probability of obtaining the measurement outcome $x_i$ given the parameter value $\theta$.

In quantum metrology, the probability distribution depends not only on the parameter-dependent quantum state, but also on the choice of measurement. Maximizing the classical Fisher information over all possible positive-operator-valued measurements (POVMs) defines the quantum Fisher information (QFI), which gives the ultimate precision bound,

$$\Delta\theta \geq \frac{1}{\sqrt{F_C(\theta)}} \geq \frac{1}{\sqrt{F_Q(\theta)}}. \tag{12}$$

Therefore, the QFI quantifies the maximum sensitivity of a quantum state to a variation of the parameter $\theta$. It is defined as the variance of the symmetric logarithmic derivative (SLD) operator $\hat{L}_\theta$, namely $F_Q(\theta) = \mathrm{Tr}\left(\hat{\rho}_\theta \hat{L}_\theta^2\right)$. The SLD has zero expectation value and satisfies

$$\frac{\partial\hat{\rho}_\theta}{\partial\theta} = \frac{1}{2}\left(\hat{\rho}_\theta\hat{L}_\theta + \hat{L}_\theta\hat{\rho}_\theta\right), \tag{13}$$

For a pure state $\hat{\rho}_\theta = |\psi(\theta)\rangle\langle\psi(\theta)|$, we have

$$\frac{\partial\hat{\rho}_\theta}{\partial\theta} = |\partial_\theta\psi(\theta)\rangle\langle\psi(\theta)| + |\psi(\theta)\rangle\langle\partial_\theta\psi(\theta)|. \tag{14}$$

Comparing this expression with the defining equation of the SLD in Eq. (13), one possible construction of the SLD operator is:

$$\hat{L}_\theta = 2\left(|\partial_\theta\psi(\theta)\rangle\langle\psi(\theta)| + |\psi(\theta)\rangle\langle\partial_\theta\psi(\theta)|\right). \tag{15}$$

Further using the normalization condition that $\langle\psi(\theta)|\partial_\theta\psi(\theta)\rangle + \langle\partial_\theta\psi(\theta)|\psi(\theta)\rangle = 0$, we can obtain the QFI for a pure state explicitly as:

$$F_Q(\theta) = 4\left[\langle\partial_\theta\psi(\theta)|\partial_\theta\psi(\theta)\rangle - |\langle\partial_\theta\psi(\theta)|\psi(\theta)\rangle|^2\right]. \tag{16}$$

This expression shows that only the component of $|\partial_\theta\psi(\theta)\rangle$ orthogonal to the state $|\psi(\theta)\rangle$ contributes to the QFI, while the parallel component merely changes the overall phase and carries no measurable information about $\theta$.

We now apply this result to the macrospic spin GHZ state in the main text:

$$|\mathrm{GHZ}\rangle = \frac{1}{\sqrt{2}}\left(|+\rangle^{\otimes N} + |-\rangle^{\otimes N}\right), \tag{17}$$

where $|+\rangle$ and $|-\rangle$ are eigenstates of the single-spin operator $\hat{\sigma}_y$,

$$\hat{\sigma}_y|+\rangle = +|+\rangle, \qquad \hat{\sigma}_y|-\rangle = -|-\rangle. \tag{18}$$

Under a transverse magnetic field $B_y$, the GHZ state evolves according to $\hat{H}_B = \gamma B_y \sum_{i=1}^{N} \hat{\sigma}_y^{(i)}$. As a result, the two branches of the superposition accumulate opposite phases over an interrogation time $t$, and the state becomes

$$|\psi(B_y)\rangle = \frac{1}{\sqrt{2}}\left(e^{2i\gamma B_y N t}|+\rangle^{\otimes N} + |-\rangle^{\otimes N}\right), \tag{19}$$

In this case,

$$F_Q(B_y) = 4\gamma^2 N^2 t^2, \tag{20}$$

and the quantum Cramér–Rao bound becomes

$$\Delta B_y \geq \frac{1}{2\gamma N t}. \tag{21}$$

The resulting sensitivity exhibits Heisenberg scaling, $\Delta B_y \propto 1/N$, in contrast to the standard quantum limit $\Delta B_y \propto 1/\sqrt{N}$ for uncorrelated spins. Physically, each spin gains the same field-dependent phase, so that in the GHZ state these phases add coherently across all $N$ spins. This produces an $N$-fold enhancement of the relative phase between the two superposition branches, and therefore a quadratic enhancement of the quantum Fisher information.

## 4. FREQUENCY OF INTERFERENCE FRINGES

In this section, we analyze the interference fringes that emerge after the two orientational branches of the macroscopic GHZ state are mechanically recombined. To quantify the frequency of the fringes, we numerically evaluate the Fourier spectrum of the probability distribution $p(\phi_s)$ and extract its dominant frequency component. The results are presented in Fig. S2. We find that the fringe frequency increases linearly with the spin number $N$, which is a direct signature of the GHZ-state structure. As $N$ increases, the oscillations become faster and finer, enabling higher phase resolution. This enhanced phase sensitivity underlies the Heisenberg-limited scaling, with $\Delta\omega \propto 1/N$.

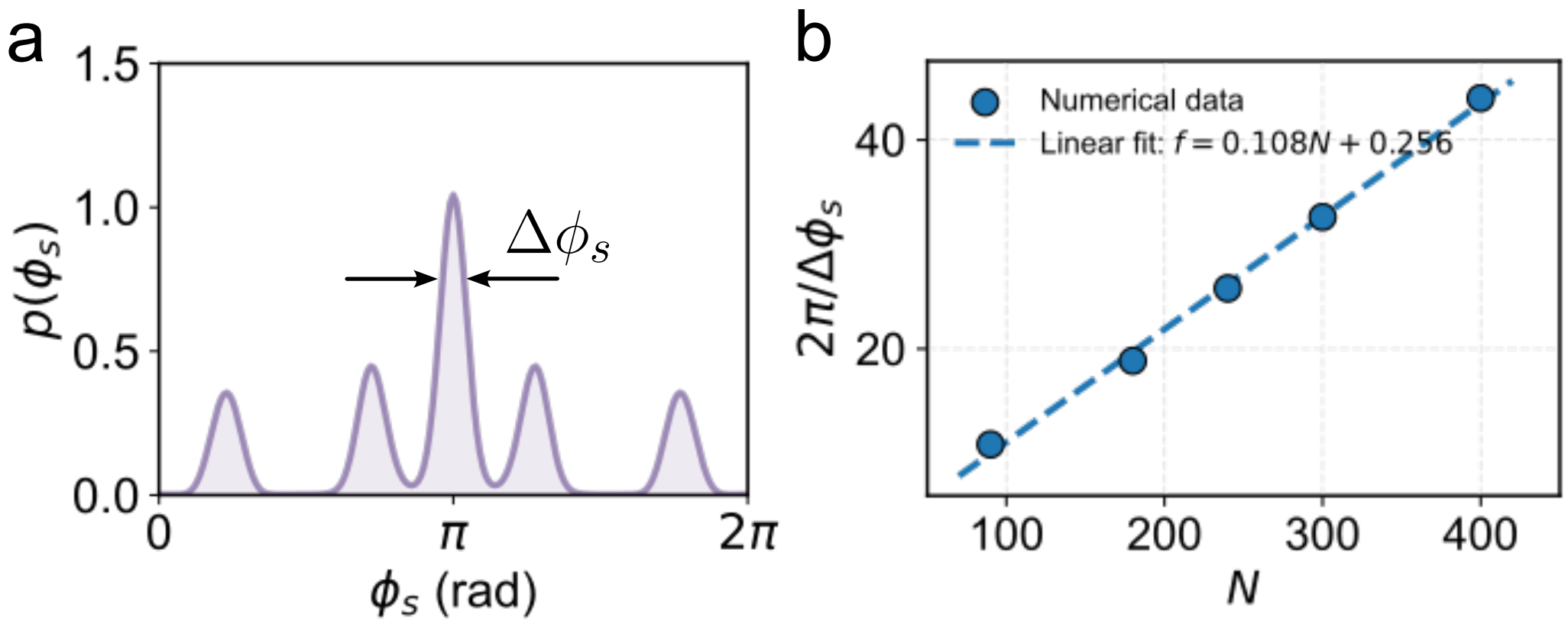


FIG. S2: (a) Illustration of the interference fringes according to the probability distribution of spin orientation $p(\phi_s)$. (b) Frequency of the fringes $2\pi/\Delta\phi_s$ across different numbers of spins $N$, showing a linear dependency.

## 5. EFFECTS OF INITIAL THERMAL FLUCTUATIONS

The generation of macroscopic superposition is often discussed in the context of ground-state cooling of the motional degree of freedom. For rotational motion, however, the relevant requirement is less stringent. Since the orientation angle is a periodic variable defined modulo $2\pi$, thermal fluctuations do not by themselves destroy the possibility of preparing a well-defined initial state, provided that the thermal angular spread remains much smaller than the full angular period, i.e.,

$$\Delta\phi \ll 2\pi. \tag{22}$$

In this regime, the rotor is still well localized, and the subsequent nonequilibrium spreading induced by the unstable dynamics can be clearly distinguished from the initial thermal width.

We therefore estimate the thermal fluctuations of a levitated rotor confined in a single-well potential centered at $\phi = 0$. Near the trapping minimum, the Hamiltonian can be approximated as

$$H = \frac{\hat{L}_z^2}{2I} + \frac{1}{2}K_\phi \hat{\phi}^2, \tag{23}$$

where $I$ is the moment of inertia and $K_\phi$ denotes the angular spring constant. The corresponding thermal state is the Gibbs state

$$\hat{\rho} = \frac{e^{-\beta H}}{Z}, \qquad \beta = \frac{1}{k_B T}, \tag{24}$$

with partition function $Z = \mathrm{Tr}\left(e^{-\beta H}\right)$.

The fluctuations are determined by the quantum expectation values

$$\langle \hat{\phi}^2 \rangle = \mathrm{Tr}(\hat{\rho}\, \hat{\phi}^2), \qquad \langle \hat{L}_z^2 \rangle = \mathrm{Tr}(\hat{\rho}\, \hat{L}_z^2). \tag{25}$$

For the thermal state of this quadratic Hamiltonian, we obtain

$$\begin{aligned} \langle \hat{\phi}^2 \rangle &= \frac{\hbar}{2\sqrt{IK_\phi}} \coth\left( \frac{\beta\hbar}{2}\sqrt{\frac{K_\phi}{I}} \right), \\ \langle \hat{L}_z^2 \rangle &= \frac{\hbar\sqrt{IK_\phi}}{2} \coth\left( \frac{\beta\hbar}{2}\sqrt{\frac{K_\phi}{I}} \right). \end{aligned} \tag{26}$$

Accordingly, the initial uncertainties of $\hat{\phi}$ and $\hat{L}_z$ are

$$\begin{aligned} \Delta\phi &= \sqrt{\frac{\hbar}{2\sqrt{IK_\phi}} \coth\left( \frac{\beta\hbar}{2}\sqrt{\frac{K_\phi}{I}} \right)}, \\ \Delta L_z &= \sqrt{\frac{\hbar\sqrt{IK_\phi}}{2} \coth\left( \frac{\beta\hbar}{2}\sqrt{\frac{K_\phi}{I}} \right)}. \end{aligned} \tag{27}$$

In the high temperature regime $k_B T > \hbar\sqrt{K_\phi/I}$, they can be approximated as the thermal fluctuations

$$\begin{aligned} \Delta\phi_{\mathrm{th}} &= \sqrt{\langle \phi^2 \rangle} = \sqrt{\frac{k_B T}{K_\phi}}, \\ \Delta L_{z,\mathrm{th}} &= \sqrt{\langle L_z^2 \rangle} = \sqrt{I k_B T}. \end{aligned} \tag{28}$$

These results show that the initial angular spread decreases with increasing angular spring constant and increases only as $\sqrt{T}$.

As an order-of-magnitude estimate, we take the experimental trapping potential reported in Ref. [2] as an example. The curvature of the translational trapping potential corresponds to an effective spring constant of order $k \sim 1$ N/m.

For a particle with radius 1 $\mu$m, the translational confinement can be mapped to an effective angular spring constant using the geometric relation $x \simeq R\phi$

$$K_\phi = kR^2 \approx 1 \times 10^{-12} \text{ J/rad}^2. \tag{29}$$

The corresponding thermal angular fluctuation at $T = 1$ K is

$$\Delta\phi_{\text{th}} \approx \sqrt{\frac{k_B T}{K_\phi}} \approx 2.6 \times 10^{-6} \text{ rad}, \tag{30}$$

which is far smaller than unity. Therefore, ground-state cooling is not strictly necessary for initializing the rotor in a well-defined orientation.

## 6. CLASSICAL SPIN-LATTICE DYNAMICS

In this section, we derive the classical equations of motion corresponding to the collective spin-lattice dynamics. The classical Hamiltonian takes the same form as the quantum one after replacing operators by classical variables,

$$H = \frac{L_z^2}{2I} - C(\mathbf{S}\cdot\mathbf{n})^2 + V(\phi), \tag{31}$$

where $\mathbf{S}\cdot\mathbf{n} = S_x \cos\phi + S_y \sin\phi$. The resulting equations of motion are

$$\frac{d\phi}{dt} = \frac{L_z}{I}, \tag{32}$$

$$\frac{dL_z}{dt} = -C\left[(S_x^2 - S_y^2)\sin 2\phi - 2S_x S_y \cos 2\phi\right] - \frac{\partial V(\phi)}{\partial\phi}, \tag{33}$$

$$\frac{dS_x}{dt} = -C\left[2S_x S_z \sin\phi\cos\phi + 2S_y S_z \sin^2\phi\right], \tag{34}$$

$$\frac{dS_y}{dt} = C\left[2S_x S_z \cos^2\phi + 2S_y S_z \sin\phi\cos\phi\right], \tag{35}$$

$$\frac{dS_z}{dt} = C\left[(S_x^2 - S_y^2)\sin 2\phi - 2S_x S_y \cos 2\phi\right]. \tag{36}$$

Using these equations, we numerically simulate the classical dynamics with a fourth-order Runge-Kutta (RK4) method. To quantify the spin-lattice locking for different spin numbers, in analogy with the corresponding quantum expectation value, we sample an ensemble of classical initial conditions and evaluate the trajectory-averaged locking parameter,

$$\langle \mathbf{s}\cdot\mathbf{n}\rangle_{\text{cl}} = \left\langle S_x\cos\phi + S_y\sin\phi\right\rangle_{\text{traj}}, \tag{37}$$

where the average is taken over all sampled classical trajectories with their associated statistical weights.

For an initial spin coherent state polarized along the $x$ direction, the collective spin is sampled from a Gaussian distribution in the transverse directions,

$$p(S_y, S_z) = \frac{1}{2\pi N}\exp\left(-\frac{S_y^2 + S_z^2}{2N}\right), \tag{38}$$

where the transverse fluctuations satisfy $\Delta S_y = \Delta S_z \propto \sqrt{N}$. The longitudinal component is then fixed by the spin-length constraint,

$$S_x = \sqrt{S_0^2 - S_y^2 - S_z^2}. \tag{39}$$

The rotor degrees of freedom are sampled from a Gaussian distribution as well,

$$p(L_z, \phi) = \frac{1}{2\pi\sigma_{L_z}\sigma_\phi} \exp\left[-\frac{L_z^2}{2\sigma_{L_z}^2} - \frac{\phi^2}{2\sigma_\phi^2}\right]. \tag{40}$$

Here, $\sigma_{L_z}$ and $\sigma_\phi$ characterize the widths of the initial angular-momentum and angle fluctuations, respectively. In the simulations, each sampled trajectory evolves independently, and the strength of spin-latice locking is obtained from the weighted ensemble average over all trajectories.

Fig. S3 shows the corresponding classical results for $N = 100$ to $N = 50000$. For a finite number of spins, perfect spin-lattice alignment is not reached, and the locking parameter remains below unity, $\langle \mathbf{s} \cdot \mathbf{n} \rangle < 1$. This deviation originates from two sources. First, the initial transverse fluctuations of a spin-coherent state scale as $1/\sqrt{N}$ after normalization, so the initial misalignment is reduced for larger $N$. Second, the dynamical fluctuations induced by the coupled spin-lattice evolution are likewise suppressed as $N$ increases and $C(\mathbf{S} \cdot \mathbf{n})^2$ dominates over $L_z^2/2I_z$. As a result, the classical trajectories become more tightly locked, and the asymptotic limit $\langle \mathbf{s} \cdot \mathbf{n} \rangle \to 1$ emerges as $N \to \infty$.

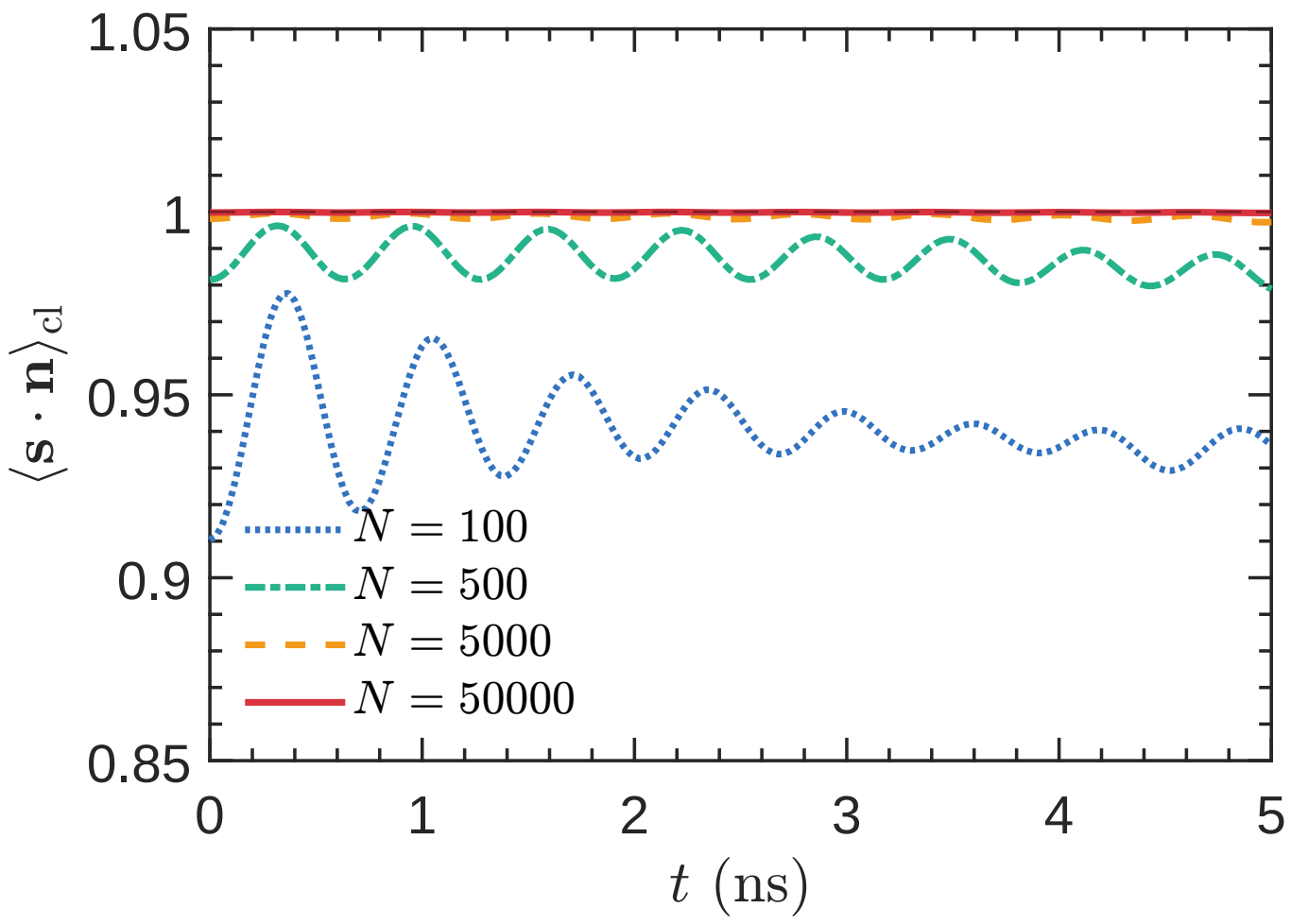


FIG. S3: Classical spin-lattice locking from ensemble-averaged trajectories for spin numbers ranging from $N = 100$ to $N = 50000$. For finite $N$, residual initial and dynamical fluctuations prevent perfect alignment, leading to $\langle \mathbf{s} \cdot \mathbf{n} \rangle_{\rm cl} < 1$. As $N$ increases, these fluctuations are suppressed, and the classical spin-lattice locking approaches the asymptotic limit $\langle \mathbf{s} \cdot \mathbf{n} \rangle_{\rm cl} \to 1$.

## 7. MASTER EQUATION FOR A QUANTUM ROTOR

To derive the master equation for the rotor, we model the planar rotor as a rigid $N$-particle assembly with masses $m_n$. In the body frame, the particles occupy fixed positions

$$\mathbf{r}_n^{(0)} = x_n \mathbf{e}_x, \tag{41}$$

subject to the center-of-mass condition $\sum_{n=1}^N m_n x_n = 0$.

If the rotor is restricted to the $x$-$y$ plane as we considered in the main text, its orientation is fully specified by a single angle $\phi$, and the position operator of the $n$-th particle becomes

$$\mathbf{r}_n = x_n \mathbf{e}_r(\hat{\phi}), \tag{42}$$

where the unit vectors are defined as

$$\mathbf{e}_r(\phi) = \mathbf{e}_x \cos\phi + \mathbf{e}_y \sin\phi, \tag{43}$$

and

$$\mathbf{e}_\phi(\phi) = \mathbf{e}_z \times \mathbf{e}_r(\phi) = -\mathbf{e}_x \sin\phi + \mathbf{e}_y \cos\phi, \tag{44}$$

with $\mathbf{e}_x$, $\mathbf{e}_y$, and $\mathbf{e}_z$ denoting the Cartesian unit vectors.

Following the procedure in Ref. [3], we begin by modeling the dynamics of the constituent point particles through the Lindblad-completed high-temperature Caldeira–Leggett master equation :

$$\partial_t \hat{\rho} = -\frac{i}{\hbar}[\hat{H}, \hat{\rho}] - \frac{i}{2\hbar}\sum_{n=1}^{N} \gamma_n [\mathbf{r}_n, \{\mathbf{p}_n, \hat{\rho}\}] - \frac{2k_B T}{\hbar^2}\sum_{n=1}^{N} m_n \gamma_n [\mathbf{r}_n, [\mathbf{r}_n, \hat{\rho}]] - \frac{1}{8k_B T}\sum_{n=1}^{N} \frac{\gamma_n}{m_n} [\mathbf{p}_n, [\mathbf{p}_n, \hat{\rho}]] . \tag{45}$$

Here the commutators and anticommutators involving vectors are defined as sums over Cartesian components, namely,

$$[\mathbf{r}_n, \{\mathbf{p}_n, \hat{\rho}\}] \equiv \sum_{j=x,y,z} [r_{n,j}, \{p_{n,j}, \hat{\rho}\}] , \quad [\mathbf{r}_n, [\mathbf{r}_n, \hat{\rho}]] \equiv \sum_{j=x,y,z} [r_{n,j}, [r_{n,j}, \hat{\rho}]] , \quad [\mathbf{p}_n, [\mathbf{p}_n, \hat{\rho}]] \equiv \sum_{j=x,y,z} [p_{n,j}, [p_{n,j}, \hat{\rho}]] . \tag{46}$$

Equivalently, this master equation may be written in the Lindblad-completed form

$$\partial_t \hat{\rho} = -\frac{i}{\hbar}\left[\hat{H} + \frac{1}{2}\sum_{n=1}^{N} \gamma_n \left(\mathbf{r}_n \cdot \mathbf{p}_n + (\mathbf{r}_n \cdot \mathbf{p}_n)^\dagger\right), \hat{\rho}\right] + \frac{2k_B T}{\hbar^2}\sum_{n=1}^{N} m_n \gamma_n \left(\mathbf{L}_n \cdot \hat{\rho}\, \mathbf{L}_n^\dagger - \frac{1}{2}\left\{\mathbf{L}_n^\dagger \cdot \mathbf{L}_n, \hat{\rho}\right\}\right), \tag{47}$$

with

$$\mathbf{L}_n = \mathbf{r}_n + \frac{i\hbar}{4m_n k_B T}\mathbf{p}_n. \tag{48}$$

We now apply the rigid-body constraint to the constituent particles. Since the rotor is confined to the $x$-$y$ plane and fully described by the single angle $\phi$, the motion of each particle is determined by the rigid rotation of the body. The velocity of the $n$-th particle is therefore

$$\dot{\mathbf{r}}_n = x_n \dot{\phi}\, \mathbf{e}_\phi(\hat{\phi}). \tag{49}$$

Using the classical rigid-body relation $L_z = I\dot{\phi}$, we replace the momentum of each particle by the corresponding operator-valued rigid-body expression

$$\mathbf{p}_n = m_n \dot{\mathbf{r}}_n = \frac{m_n x_n}{I}\, \mathbf{e}_\phi(\hat{\phi})\hat{L}_z. \tag{50}$$

Substituting Eqs. (42) and (50) into the Hamiltonian correction term in Eq. (47), we find

$$\begin{aligned} \mathbf{r}_n \cdot \mathbf{p}_n &= x_n \mathbf{e}_r(\hat{\phi}) \cdot \frac{m_n x_n}{I}\, \mathbf{e}_\phi(\hat{\phi})\hat{L}_z \\ &= \frac{m_n x_n^2}{I}\left(\mathbf{e}_r(\hat{\phi}) \cdot \mathbf{e}_\phi(\hat{\phi})\right)\hat{L}_z = 0, \end{aligned} \tag{51}$$

since

$$\mathbf{e}_r(\hat{\phi}) \cdot \mathbf{e}_\phi(\hat{\phi}) = 0. \tag{52}$$

Hence the Hamiltonian correction vanishes.

The Lindblad operator for the $n$-th particle then becomes

$$\begin{aligned} \mathbf{L}_n &= x_n \mathbf{e}_r(\hat{\phi}) + \frac{i\hbar}{4m_n k_B T}\frac{m_n x_n}{I}\, \mathbf{e}_\phi(\hat{\phi})\hat{L}_z \\ &= x_n \left[\mathbf{e}_r(\hat{\phi}) + \frac{i\hbar}{4k_B T I}\, \mathbf{e}_\phi(\hat{\phi})\hat{L}_z\right]. \end{aligned} \tag{53}$$

Defining

$$\mathbf{A} = \mathbf{e}_r(\hat{\phi}) + \frac{i\hbar}{4k_B T I}\mathbf{e}_\phi(\hat{\phi})\hat{L}_z, \tag{54}$$

we have simply

$$\mathbf{L}_n = x_n \mathbf{A}. \tag{55}$$

Substituting this into Eq. (47) yields

$$\partial_t \hat{\rho} = -\frac{i}{\hbar}[\hat{H}, \hat{\rho}] + \frac{2D_\phi}{\hbar^2}\left(\mathbf{A}\cdot\hat{\rho}\,\mathbf{A}^\dagger - \frac{1}{2}\left\{\mathbf{A}^\dagger\cdot\mathbf{A}, \hat{\rho}\right\}\right), \tag{56}$$

where

$$D_\phi = k_B T \sum_{n=1}^{N} m_n \gamma_n x_n^2. \tag{57}$$

Using the above dissipator, we now estimate the orientational decoherence rate. In the angle representation,

$$\rho(\phi, \phi') = \langle \phi | \hat{\rho} | \phi' \rangle, \tag{58}$$

the dominant contribution in the high-temperature regime comes from the position-dependent part of the Lindblad operator,

$$\mathbf{A} \simeq \mathbf{e}_r(\hat{\phi}), \tag{59}$$

since the momentum-dependent correction is suppressed by the factor $\hbar/(4k_B T I)$. The dissipator therefore reduces approximately to

$$\mathcal{D}\hat{\rho} \simeq \frac{2D_\phi}{\hbar^2}\left(\mathbf{e}_r(\hat{\phi})\cdot\hat{\rho}\,\mathbf{e}_r(\hat{\phi}) - \hat{\rho}\right). \tag{60}$$

Taking the matrix element between angle eigenstates yields

$$\langle \phi | \mathcal{D}\hat{\rho} | \phi' \rangle \simeq \frac{2D_\phi}{\hbar^2}\left(\mathbf{e}_r(\phi)\cdot\mathbf{e}_r(\phi') - 1\right)\rho(\phi, \phi'). \tag{61}$$

Since

$$\mathbf{e}_r(\phi)\cdot\mathbf{e}_r(\phi') = \cos(\phi - \phi'), \tag{62}$$

we obtain

$$\partial_t \rho(\phi, \phi') \simeq -\frac{2D_\phi}{\hbar^2}\left[1 - \cos(\phi - \phi')\right]\rho(\phi, \phi'). \tag{63}$$

Therefore, the orientational coherence between two angles $\phi$ and $\phi'$ decays exponentially with rate

$$\Gamma_{\rm dec}(\phi, \phi') = \frac{2D_\phi}{\hbar^2}\left[1 - \cos(\phi - \phi')\right]. \tag{64}$$

Equivalently, using

$$1 - \cos(\phi - \phi') = 2\sin^2\left(\frac{\phi - \phi'}{2}\right), \tag{65}$$

this can be written as

$$\Gamma_{\rm dec}(\phi, \phi') = \frac{4D_\phi}{\hbar^2}\sin^2\left(\frac{\phi - \phi'}{2}\right). \tag{66}$$

The corresponding decoherence time is thus estimated as

$$T_{\rm dec}(\phi, \phi') \sim \frac{\hbar^2}{2D_\phi\left[1 - \cos(\phi - \phi')\right]}. \tag{67}$$

For small angular separations $|\phi - \phi'| \ll 1$, we have

$$1 - \cos(\phi - \phi') \simeq \frac{(\phi - \phi')^2}{2}, \tag{68}$$

so that

$$\Gamma_{\rm dec}(\phi, \phi') \simeq \frac{D_\phi}{\hbar^2}(\phi - \phi')^2. \tag{69}$$

To obtain the explicit pressure and temperature dependence of $D_\phi$, we now adopt a classical description of gas collisions. This is justified in the dilute high-temperature regime considered here, where the gas acts as a Markovian thermal bath, and the rotational damping can be calculated from classical free-molecular kinetics. The diffusion constant follows from the fluctuation-dissipation relation

$$D_\phi = k_B T I \eta_\phi, \tag{70}$$

where $\eta_\phi$ denotes the effective rotational damping coefficient, defined via the damping torque

$$N_z = -I\eta_\phi\omega. \tag{71}$$

Now consider a cylinder rotor of radius $R$ and length $L$ rotating about its symmetry axis with angular velocity $\omega$. To derive the tangential force and the resulting torque, we introduce at each surface element an instantaneous local frame $(\mathbf{e}_n, \mathbf{e}_t, \mathbf{e}_u)$, where $\mathbf{e}_n$ denotes the outward normal, $\mathbf{e}_t$ the local tangential direction of the surface motion, and $\mathbf{e}_u$ completes the orthonormal triad. In this frame, the surface element moves with tangential velocity

$$\boldsymbol{v}_{\rm surf} = v_\parallel \mathbf{e}_t. \tag{72}$$

The surrounding gas is Maxwell distributed in the laboratory frame. In the instantaneous frame of the surface element, the distribution is shifted along the tangential direction,

$$f(\mathbf{v}) = n\left(\frac{m}{2\pi k_B T}\right)^{3/2} \exp\left[-\frac{m\left(v_n^2 + v_u^2 + (v_t + v_\parallel)^2\right)}{2k_B T}\right], \tag{73}$$

where $n$ is the number density, $m$ the molecular mass, and $(v_n, v_u, v_t)$ the velocity components in the local frame. For $v_\parallel \ll \sqrt{k_B T/m}$, expansion to first order in $v_\parallel$ yields

$$f(\mathbf{v}) \simeq f_0(\mathbf{v}) - \frac{m v_\parallel v_t}{k_B T} f_0(\mathbf{v}), \tag{74}$$

with the equilibrium distribution

$$f_0(\mathbf{v}) = n\left(\frac{m}{2\pi k_B T}\right)^{3/2} \exp\left[-\frac{m(v_n^2 + v_u^2 + v_t^2)}{2k_B T}\right]. \tag{75}$$

The tangential force on the surface element is determined by the incoming tangential momentum flux,

$$dF_\parallel = dA \int_{v_n<0} (-v_n)\,(m v_t)\, f(\mathbf{v})\, d^3v. \tag{76}$$

The zeroth-order term vanishes by symmetry, while the first-order contribution gives

$$dF_\parallel = -\frac{m^2 v_\parallel}{k_B T}\, dA \int_{v_n<0} (-v_n) v_t^2\, f_0(\mathbf{v})\, d^3v. \tag{77}$$

Evaluating the Gaussian integral, we obtain

$$dF_\parallel = -\frac{n}{2}\sqrt{\frac{2mk_BT}{\pi}}\, v_\parallel dA = -\frac{P}{2}\sqrt{\frac{2m}{\pi k_B T}}\, v_\parallel dA, \tag{78}$$

where $P = nk_BT$ is the gas pressure.

For the cylindrical side surface, we have $v_\parallel = \omega R$ and $r_\perp = R$, yielding

$$N_z^{(\text{side})} = \int_{\text{side}} r_\perp \, dF_\parallel = -P\pi R^3 L\sqrt{\frac{2m}{\pi k_B T}}\,\omega. \tag{79}$$

For each circular end cap, the local tangential velocity is $v_\parallel = \omega r$, so that

$$N_z^{(\text{cap})} = \int_{\text{cap}} r dF_\parallel = -\frac{P}{2}\sqrt{\frac{2m}{\pi k_B T}}\omega\frac{\pi R^4}{2}. \tag{80}$$

With two end caps, the total contribution becomes

$$N_z^{(\text{ends})} = -\frac{P}{2}\sqrt{\frac{2m}{\pi k_B T}}\omega\,\pi R^4. \tag{81}$$

The total damping torque is therefore

$$N_z = N_z^{(\text{side})} + N_z^{(\text{ends})} = -\frac{P}{2}\sqrt{\frac{2\pi m}{k_B T}}R^3 L\left(2 + \frac{R}{L}\right)\omega. \tag{82}$$

Substituting the above expression into Eq. (71) and using the fluctuation–dissipation relation Eq. (70), we obtain

$$\eta_\phi = \frac{PR^3L}{2I}\sqrt{\frac{2\pi m}{k_B T}}\left(2 + \frac{R}{L}\right), \qquad D_\phi = \frac{1}{2}PR^3L\sqrt{2\pi m k_B T}\left(2 + \frac{R}{L}\right). \tag{83}$$

## 8. MAGNETIC NOISE AND SPIN DEPHASING

A macrospin GHZ state is far more fragile to magnetic noise than a single spin, which is the direct trade-off of its high phase sensitivity. In experiments, maintaining a macrospin GHZ state therefore requires the ambient magnetic field noise to be sufficiently suppressed through magnetic shielding, and this requirement may ultimately set the practical limit on the achievable superposition size $N$.

Consider magnetic field noise $\delta B_y(t)$ coupled to the macrospin through the Zeeman interaction,

$$H_B(t) = -\gamma\,\delta B_y(t)\,\hat{S}_y. \tag{84}$$

The fluctuating field produces a random relative phase between the two branches $|+\rangle^{\otimes N}$ and $|-\rangle^{\otimes N}$,

$$\delta\phi(t) = 2\gamma N \int_0^t \delta B_y(t')\,dt'. \tag{85}$$

The off-diagonal element of the density matrix therefore has a stochastically fluctuating phase,

$$\rho_{+-}(t) = \rho_{+-}(0)\,e^{i\,\delta\phi(t)}. \tag{86}$$

The spin dephasing reflects the ensemble average over many different noises, where the off-diagonal term eventually decays.

We consider the limiting case of Gaussian white noise, where the fluctuation $\delta\phi$ has zero mean and a Gaussian distribution. The power spectral density of the magnetic field noise is flat with no temporal memory, $\langle\delta B_y(t_1)\delta B_y(t_2)\rangle = S_B\delta(t_1 - t_2)$, where $S_B$ is the noise power spectral density. In this case, the variance of the phase factor over different

stochastic trajectories is:

$$\begin{aligned}\langle \delta\phi^2(t) \rangle &= (2\gamma N)^2 \int_0^t \int_0^t \langle \delta B_y(t_1)\, \delta B_y(t_2) \rangle \, dt_1\, dt_2 \\ &= (2\gamma N)^2 S_B \int_0^t \int_0^t \delta(t_1 - t_2)\, dt_1\, dt_2 \\ &= (2\gamma N)^2 S_B\, t. \end{aligned} \tag{87}$$

For a Gaussian random variable, the ensemble average of the phase factor follows from the characteristic function of a Gaussian distribution,

$$\langle e^{i\delta\phi} \rangle = e^{-\langle \delta\phi^2 \rangle /2}, \tag{88}$$

so the ensemble-averaged off-diagonal element decays as

$$\langle \rho_{+-}(t) \rangle = \rho_{+-}(0) \exp\!\left( -\frac{(2\gamma N)^2 S_B}{2}\, t \right). \tag{89}$$

Thus, the dephasing rate of a spin GHZ state is $\Gamma_\phi^{(N)} = (2\gamma N)^2 S_B/2$, which is enhanced as $N^2$ relative to a single spin.

To maintain GHZ coherence over a time window $t$, we require $\Gamma_\phi^{(N)} t \lesssim 1$. In terms of the noise amplitude spectral density $\sqrt{S_B}$,

$$\sqrt{S_B} \lesssim \frac{1}{\sqrt{2}\gamma N \sqrt{t}}. \tag{90}$$

For an effective spin number $N = 10^7$ and time window of 1 $\mu s$, the required noise amplitude spectral density needs to be below 0.4 fT/$\sqrt{\mathrm{Hz}}$. Meeting this bound requires advanced magnetic shielding, along with the suppression of thermal magnetic noise and eddy-current-induced fluctuations.


\fi

\end{document}